\def\year{2016}
\documentclass[letterpaper]{article}

\usepackage{aaai}
\usepackage{times}
\usepackage{helvet}
\usepackage{courier}

\usepackage[hidelinks]{hyperref}

\usepackage{booktabs}
\usepackage{natbib}
\usepackage{subfigure}
\usepackage{graphicx}
\usepackage{amsmath,amsfonts}
\usepackage{flushend}
\usepackage{comment}
\usepackage[usenames,dvipsnames,table]{xcolor}

\newcommand{\todo}[2]{}

\nocopyright

\begin{document}

\title{Hot-Rodding the Browser Engine: Automatic Configuration of JavaScript Compilers}

\author{Chris Fawcett \and Lars Kotthoff \and Holger~H. Hoos \\
    University of British Columbia \\
    Department of Computer Science \\
    \{fawcettc, larsko, hoos\}@cs.ubc.ca
}

\maketitle

\begin{abstract}
Modern software systems in many application areas offer to the user a multitude of parameters,
switches and other customisation hooks. Humans tend to have difficulties determining the best
configurations for particular applications. Modern optimising compilers are an example of such
software systems; their many parameters need to be tuned for optimal performance,
but are often left at the default values for convenience.

In this work, we automatically determine compiler parameter settings that
result in optimised performance for particular applications.
Specifically, we apply a state-of-the-art automated parameter configuration procedure
based on cutting-edge machine learning and optimisation techniques to two prominent
JavaScript compilers and demonstrate that significant performance improvements,
more than 35\% in some cases, can be achieved over the default parameter settings on a diverse set of benchmarks.
\end{abstract}

\section{Introduction}

JavaScript is one of the fundamental technologies underpinning the world wide
web today. From its humble beginnings as a scripting language to support basic
interactive content, it has matured to the point where it powers large applications for
multi-billion dollar businesses. In addition to client-side JavaScript run in
the user's browser, server-side JavaScript is becoming increasingly popular for
high-throughput, low-latency web applications.

The JavaScript language (actually ECMAScript) is one of the most popular languages
today\footnote{\url{https://github.com/blog/2047-language-trends-on-github}}.
The increasing complexity of JavaScript applications and the deployment in
environments with high performance requirements has driven the development of
JavaScript compilers that produce more efficient, highly-optimised code. These
compilers are complex pieces of software themselves and make a plethora of
configurable parameters available to the user -- one size does not fit all, and
how exactly a piece of code should be optimised may depend on the particular
application and execution environment.

While many JavaScript code optimisers exist, these usually focus
on ``compressing'' the source code so that it can be transferred from the
server to the browser more efficiently by means of source-level transformations.
Such optimisations do not affect the semantics of the code nor how it is compiled, and they usually do not improve performance
in terms of running time.

There are many benefits to optimising not only the code but also the JavaScript
compiler for particular applications. On mobile devices, power consumption
is a major issue, and optimised code can help reduce it. On a server, reducing
the running time of a software component means that more transactions can be supported on
the same hardware. On a desktop machine, lag in interactions can be reduced and user
interfaces be made more responsive.

In the vast majority of applications, the JavaScript
compiler is run in its default configuration, which has been chosen by its developers to achieve robust performance
across a broad range of use cases. While these default settings will provide reasonable
performance in most situations, we demonstrate that often, substantial gains can be realised
easily, by searching the configuration space for compiler parameter settings that better optimise the
JavaScript code produced for a particular application. With little effort, applications can
be made more efficient and consume less resources. This effect even holds for the large,
heterogeneous benchmark sets used during development of the JavaScript engines themselves, upon
which the default settings are purportedly based.

Compared to traditional applications of automated algorithm configuration,
JavaScript code often runs for only relatively short periods of time, but very frequently.
The same code can be run millions of times a day, each time a website is loaded or a request is made to a
server. Even small improvements translate to massive aggregate savings in
resources. 

We apply state-of-the-art machine learning techniques for automatic parameter
configuration to the two main JavaScript compilers. On a range of popular and
representative benchmarks, we show that performance can be improved by more than
35\% even with relatively modest configuration effort, without any modification to the JavaScript source code
under consideration, or of the JavaScript engine running it, other than the change in parameter configuration.

\section{Background}

The idea of optimising the configuration of a compiler for a particular
application or set of applications is not new. The Milepost GCC
project~\cite{fursin_milepost_2011} is perhaps the most prominent example and
uses machine learning to dynamically determine the best level of optimisation.
In an iterative process, it can improve execution time, code size, compilation
time and other metrics. The approach has been integrated into the widely-used
GCC compiler. Other approaches that optimise the code generation for C programs
include \cite{haneda_automatic_2005,pan_fast_2006,plotnikov_automatic_2013}.
While most of these optimise the GCC compiler, there exists some work
on LLVM as well~\cite{fursin_collective_2014}.

Another focus of research for automatic dynamic optimisation of compiled code
has been the Jikes Java compiler~\cite{alpern_jikes_2005}.
\citet{hoste_automated_2010}~use multi-objective evolutionary search to identify
configurations that are Pareto-optimal in terms of compilation time and code
quality. \citet{cavazos_method-specific_2006}~learn logistic regression models
that predict the best optimisation to apply to a method.
\citet{kulkarni_mitigating_2012}~use artificial neural networks to determine the
order in which a set of optimisations should be applied during compilation.

A major concern with all compiler configuration optimisation approaches is the
computational effort required to determine a good or optimal configuration. If this is too
large, any benefits gained through the optimisation may be negated. One approach
to reducing the initial overhead is to move the configuration process online and
to learn to identify good configurations over successive compilations, but other
approaches have been explored in the literature
(see, e.g.~\cite{thomson_reducing_2010,ansel_siblingrivalry_2012,tartara_continuous_2013}).


Compilers that translate JavaScript to native code are relatively new compared
to compilers for more established languages likes C. While they are also highly
optimised and, in the case of JavaScriptCore through the use of the LLVM
framework, leverage at least some of the benefits decades of optimisation effort
has brought to compilers for other languages, we believe that performance
improvements over the default configuration are to be gained more easily here.
Furthermore, due to the widespread use of JavaScript in applications with hundreds of millions
of end users (such as web browsers), any performance improvements are likely to be impactful.

\subsection{JavaScript Optimisation}

Existing JavaScript optimisers, such as Google's Closure
Tools\footnote{\url{https://developers.google.com/closure/}} and Yahoo's YUI
compressor\footnote{\url{https://yui.github.io/yuicompressor/}}, focus on
source code transformations that do not alter the syntax or semantics of the
code, but compress the representation by shortening identifiers, removing
white space or inlining code. The aim of these optimisations is to reduce the
size of the code that has to be transferred from the server to the user's
browser, thereby reducing the load time of the page. It focuses on efficiency
\emph{before} the code is run, but does nothing to improve performance
\emph{while} the code is running.

Indeed, many of those tools and techniques are not specific to JavaScript, but
are also applied to other resources that are transferred to the client when a
web page is loaded, such as Cascading Stylesheets (CSS). In contrast, what we
propose here leverages the specific configuration options of JavaScript engines
to optimise the actual runtime behaviour and efficiency of the code.

One attractive aspect of our approach is that it naturally complements any
extensions implemented to an existing JavaScript engine (by performing our automated configuration procedure again),
and is able to search for improving engine configurations while consuming commodity
compute cycles, without significant impact on development and engineering effort. 
Running an automated configuration procedure on a commodity compute cluster 
for a week is significantly cheaper than the salary of even a single engineer for the same period, 
and optimising the engine configuration automatically frees up human development resources, which can then be used
to further enhance the JavaScript engine with new or improved optimisation mechanisms.

\todo{CF}{I feel as if there is a small amount more to say here regarding the suitability of JavaScript engine parameters
to automated algorithm configuration. Specifically, that we are in essence solving extremely similar (or truly identical!) problem
instances potentially hundreds of millions of times. This would be extremely strange in, say, the SAT setting. We do mention this a bit
in the introduction, but I think it should be stressed and this might be a good
spot.}

\todo{LK}{I don't think that this makes JS inherently more suitable -- we can't
expect the massive gains seen in other domains. For real-world stuff, this is
probably even worse as a large part will be IO.}

\subsection{Automated Algorithm Configuration}

Most software has switches, flags and options through which the user can
control how it operates. As the software becomes more complex or is used
to solve more challenging and diverse problems, the number of these options also
tends to increase. While some of these parameters control the input/output behaviour 
of a given piece of software or algorithm, others merely affect efficiency in terms of 
resource use.

The algorithm configuration problem is concerned with finding the best parameter
configuration for a given algorithm on a set of inputs, where the
definition of ``best'' can vary, depending on the given application scenario. In many practical cases, the
goal is to achieve better performance, and this is how we use algorithm configuration here  -- we want to achieve the same
functionality, but with reduced resource requirements. Specifically, in this work
we focus on minimizing the CPU time required, but in principle, any scalar measure
of performance can be used.

Finding the best parameter configuration for a given algorithm is a long-standing
problem. Humans tend to be bad at solving it -- evaluating parameter configurations
requires substantial effort, and interactions between parameters may be complex and unintuitive.
\citet{minton_automatically_1996} notes that,
\begin{quote}
``Unlike our human subjects, [the system] experimented with a wide variety of
combinations of heuristics. Our human subjects rarely had the inclination or
patience to try many alternatives, and on at least one occasion incorrectly
evaluated alternatives that they did try.''
\end{quote}
Fortunately, there exist many automated procedures for algorithm configuration.
Perhaps the simplest approach is to try all combinations of parameter values.
This approach is known as a full factorial design in the statistics literature on experimental design
and as grid search in computer science (specifically, in machine learning); its main
disadvantage lies in its high cost -- the number of configurations to be evaluated grows
exponentially with the number of parameters and their values. For most practical
applications, including the ones we consider in the following, complete grid search is infeasible.

A commonly used alternative is simple random sampling:
Instead of evaluating every combination of parameter values, we randomly sample
a small subset. This is much cheaper in practice and achieves surprisingly good
results~\cite{bergstra_random_2012}. 
Indeed, in machine learning, random sampling is a widely used method for hyper-parameter optimisation.
Unfortunately, when searching high-dimensional configuration spaces, random sampling is known
to achieve poor coverage and can waste substantial effort evaluating poorly performing 
candidate configurations.

A more sophisticated approach to algorithm configuration is provided by so-called racing
methods~\cite{birattari_racing_2002}, which iteratively evaluate candidate configurations
on a series of inputs and eliminate candidates as soon as they can be shown to 
significantly fall behind the current leader of this race.
Local search based configurators, on the other hand, 
iteratively improve a given configuration by applying small changes and avoid stagnation 
in local optima by means of diversification techniques (see, e.g., \cite{hutter_paramils_2009}).

More recently, model-based algorithm configuration methods have gained prominence. 
These are based on the key idea of constructing a model of how the parameters affect performance;
this empirical performance model is then used to select candidate configurations to be evaluated
and updated based on the results from those runs.
Arguably the best known model-based configurator (and the current state of the art) is
SMAC~\cite{hutter_sequential_2011}, which we use in the following.

SMAC and the other general-purpose algorithm configuration methods mentioned
above have been applied with great success to a broad range of problems,
including propositional satisfiability~\cite{hutter_automatic_2007}, mixed
integer programming~\cite{hutter_automated_2010}, machine learning
classification and regression~\cite{thornton_auto-weka_2013}, and improving the
performance of garbage collection in Java~\cite{lengauer_taming_2014}.

The existence of effective algorithm configuration procedures has implications for the
design and development of high-performance software. Namely, rather than limiting design
choices and configurable options to make it easier (for human developers) to find good settings,
there is now an incentive to introduce, expose and maintain many design choices,
and to let automated configuration procedures find performance-optimized configurations for specific application contexts.
This is the core idea behind the recent \emph{Programming by Optimization (PbO)} paradigm~\cite{pbo}.

However, if software is not developed using specific tools supporting PbO, the application of automated
configuration procedures requires the manual specification of a \emph{configuration space} based on the definitions of and constraints on all configurable parameters.
For complex and highly parameterised software, such as the JavaScript engines we consider in this work,
this process can be somewhat involved, since it not only involves collecting the names and domains 
(i.e., permissible values) for all parameters, but also conditionality relations between them (e.g., parameter $a$'s value only matters if parameter $b$ has value $x$),
and constraints that rule out certain configurations (e.g., configurations known to cause crashes or faulty behaviour).

Furthermore, in typical applications of automated algorithm configuration, developers
need to carefully construct a set of `training' inputs that is representative of those encountered
in the intended application context of the algorithm or software to be configured.
If automated configuration is applied to produce a performance-optimised configuration using training inputs unlike those
seen in typical use, the resulting configuration is unlikely to perform as well in the actual application as on the training set used
as the basis for configuration. (This, of course, also holds for manual configuration, but the effect tends to become more pronounced if more effective optimisation methods are used.)

Interestingly, JavaScript engine parameter optimisation (and, more generally, certain flavours of compiler optimisation)
differs from most other applications of automated algorithm configuration, in that it makes sense to use a training set consisting of a single input in the form of a program source, whose performance is to be optimised by means of compilation and execution with specific engine parameters.
\todo{HH}{Is it just compilation or also execution?CF: Fixed.}
Consider a popular Node.js application running a JavaScript workload that does not
change appreciably for each request it receives. Any performance increases on that particular workload
are of immediate, significant benefit, and performance decreases on other hypothetical workloads are irrelevant.
These situations are ideal for our approach, as they allow for the performance gains achieved in offline performance optimisation to be leveraged across potentially hundreds of millions of future runs of the software thus optimised.

\section{Automated Configuration of JavaScript Engines}

\subsection{JavaScript Engines}

We consider two state-of-the-art JavaScript engines in this work;
JavaScriptCore\footnote{\url{https://trac.webkit.org/wiki/JavaScriptCore}} and
Google's V8\footnote{\url{https://code.google.com/p/v8/}}.
This choice was motivated by the popularity and availability of these engines, rather than absolute performance.
We note that our goal was not to compare the performance of the two engines, but rather to investigate to what extent the default configuration of each can be improved. 

JavaScriptCore (or JSC) is an optimising JavaScript virtual machine
developed as the JavaScript engine for WebKit; it is used in Apple's Safari browser on both OS X and iOS,
as well as in many other Apple software projects, web browsers, and in a WebKit extension of Node.js.
It contains a low-level interpreter (LLInt), a simpler baseline just-in-time (JIT) compiler, another JIT compiler
optimizing for low latency (DFG JIT), and a JIT compiler optimizing for high throughput (FTL JIT). All of these components
can be active simultaneously for different blocks of code, based on execution thresholds, and blocks can be optimised (and deoptimised) between them many times. In fact, a recursive function can be executing in different JITs (or the LLInt) simultaneously at
different levels of the recursion. Our JSC parameter space contains 107
parameters (Table~\ref{tab:compiler-parameter-spaces}), where most
of the parameters have numerical domains. These numerical parameters mostly control counters and thresholds for activating
various functionality, and for triggering optimisation/deoptimisation between the LLInt and the various JITs.

\begin{table}
    \begin{center}
        \begin{tabular}{r@{\hskip 2em}r@{\hskip 1em}rrr}
                            & & \multicolumn{3}{c}{\# parameters of type} \\
\textbf{Engine} & \# parameters     & Boolean & integer & real \\
\midrule
JSC             & 107       & 40      & 54      & 13 \\
V8              & 173       & 143     & 30      & 0 \\

        \end{tabular}
        \caption{
            For each of the two JavaScript engines considered in this work, we give the
            total number of parameters in the configuration space as well as how many have
            Boolean, integer and real-valued domains, respectively. (There are no parameters
            with categorical domains in either configuration space.)
        }
        \label{tab:compiler-parameter-spaces}
    \end{center}
\end{table}

The V8 JavaScript engine was initially developed for Google's Chrome browser
and is now used in other web browsers such as Opera, in server-side applications
using projects like Node.js\footnote{\url{https://nodejs.org/}} and as a library
embedded in other software applications.
V8 is somewhat unique in that it does not contain an interpreter, but instead compiles
JavaScript code blocks directly to native machine code when they are first encountered, which is then
optimised continuously over the course of running on a given input. Our interpretation of V8's parameter configuration space contains 173 parameters, primarily
Boolean choices to enable or disable specific functionality. The remaining integer parameters
control various aspects of that functionality, including inlining levels, loop unrolling, garbage collection thresholds and
stack frame sizing.

In order to specify the parameter configuration space for our two JavaScript
engines, JSC and V8,
we determined the name and type of each parameter, based on the documentation and command-line parser source code.
Unfortunately, domains for the numerical parameters are not specified by the developers, and only few conditional dependencies are explicitly described. We therefore had to resort to educated guesses; when in doubt, we aimed to err on the side of larger domains (within reason). Each space was then refined by sampling 100\,000 random configurations and running the engines on a simple problem instance to check for segmentation faults and other abnormal behaviour. For both engines,
many crashing configurations were thus identified, leading to iterative refinement of the configuration spaces through domain reduction as well as by adding forbidden parameter combinations and conditional parameter dependencies.

\subsection{Benchmark Instances}

We have selected four benchmark sets containing heterogeneous JavaScript problem instances,
identified as relevant to the JavaScript engine development community and end
users.
We aimed to avoid bias towards benchmark sets preferred by particular
development teams. In particular, we included the benchmark sets developed by the
developers of JSC and V8.

Our benchmark suite comprises the Octane 2.0~\footnote{\url{https://developers.google.com/octane/}}, 
SunSpider 1.0.2~\footnote{\url{https://www.webkit.org/perf/sunspider/sunspider.html}},
Kraken 1.1~\footnote{\url{http://krakenbenchmark.mozilla.org}} and Ostrich~\cite{khan_using_2014} benchmark sets.
We created harnesses that allowed us to execute and measure these benchmarks
programmatically, outside of a browser environment. We note that the techniques we
use here readily extends to browser-based settings, albeit the integration effort would be higher.

Octane 2.0 is Google's JavaScript compiler benchmark suite and includes
18 real-world benchmarks that range over different types of tasks, including a 2D
physics engine, a PDF rendering engine, a portable game system emulator,
a regular expression generator as well as instances testing, e.g., node allocation and reclamation.

The SunSpider 1.0.2 benchmark set was developed by the Web\-Kit team and 
contains 26 problem instances representing
a variety of different tasks that are relevant to real-world applications,
including string manipulation, bit operations, date formatting and cryptography.

Kraken 1.1 was developed by Mozilla and contains 14 problem instances that were extracted from
real-world applications and libraries. These benchmarks primarily cover web-specific tasks (e.g., JSON parsing),
signal processing (e.g., audio and image processing), cryptography (e.g., AES, PBKDF2, and SHA256 implementations)
and general computational tasks, such as combinatorial search.

Ostrich is based on benchmark suites for important numerical computation tasks, such
as OpenDwarf~\cite{feng_opencl_2012}. While the other benchmarks focus on the
types of computations that are common on the web, Ostrich provides a way to
measure the performance on computations that are becoming increasingly relevant
as JavaScript gains in popularity and is deployed in new contexts.

\subsection{Experimental Setup}

All of the experiments reported in the following were performed using a single
Microsoft Azure Cloud instance of type ``G5'' running a standard installation
of Ubuntu 15.04. This instance type has two 16-core processors with a total of 448GB of RAM;
it is the sole user of the underlying hardware, based on a one-to-one mapping to two Intel Xeon E5-2698A v3 processors.

We use JavaScriptCore r188124 and V8 version 4.6.40, release builds compiled from source using GCC 4.9.2.
Our version of the SMAC configurator is v2.10.03\footnote{\url{http://www.cs.ubc.ca/labs/beta/Projects/SMAC/}}, run using Oracle Java JDK 1.8.0\_25.

For each of our configuration scenarios, we performed 25 independent runs of SMAC with a 1 CPU-day runtime cutoff,
allocating a maximum of 60 CPU seconds to each run on a particular problem instance. The objective value to be minimised by
SMAC is the so-called \emph{Penalised Average Runtime} (PAR) score, which penalises timed-out and crashing runs by assigning them
an objective value of 10 times the runtime cutoff (PAR-10), and otherwise assigns an objective value of the CPU time used.
This greatly disincentivizes bad and invalid configurations, in order to bias the configurator against selecting them.

The incumbent configuration with the best PAR-10 score reported by SMAC after termination was selected as the final result of the configuration process, and a subsequent validation
phase was performed to run both the JSC/V8 default configuration and the optimised configuration selected by our procedure on the
entire problem instance set. For these validation runs, we perform 100 runs per configuration and benchmark instance, and compute the PAR-10 score across all runs for each configuration.
\todo{HH}{Not PAR-10? CF: How's this?}

We require repeated runs to obtain statistically stable results. Individual runs are very
short and subject to susbstantial noise from the environment, e.g.\ operating system
jobs and contention for shared memory. Through repeated runs and averaging, we
achieve more realistic results that are less affected by very short and very
long outlier runs.

\section{Empirical Results}

The purpose of our experiments is twofold. First, we intend to demonstrate that
the performance \emph{across a set of diverse benchmarks} can be improved by
using a different parameter configuration than the default. This would indicate that
compiler developers may want to adjust the default settings with which they ship
their compilers, or that users that focus on particular types of applications may wish to
do so themselves. It also demonstrates the potential for techniques that periodically adjust 
the configuration of the JavaScript engine based on the types of JavaScript code run recently.

The second part of our experiments focusses on \emph{specific individual
benchmarks} and shows that performance can be improved significantly by
specialising the compiler configurations to a specific piece of code to run, rather
than being forced to accept tradeoffs due to competing requirements by a heterogeneous set of benchmarks.
This finding can be exploited in two ways: Users who run the same piece of JavaScript code
over and over again (e.g., in a server-side JavaScript application) 
can benefit from offline tuning, while at the same time, very
short online configuration runs for code that a user's browser accesses
frequently can potentially optimise its performance.

\todo{LK}{Can we show some
examples of how many iterations SMAC needs before it finds a significantly
better than default config to give some credence to this? Maybe even random
sampling? HH: Good question.}

\todo{CF}{I can still trawl through the SMAC logs if we have time.}

\subsection{Results on Benchmark Sets}

\todo{HH}{I've rearranged the following to make it sound less defensive. If you like this, please rearrange the table to show the JSC results first. We may then also want to rearrange the order in which the two engines are introduced and discussed in 3.1. I like doing this, because it makes the order lexicographic.}

As can be seen in Table~\ref{tab:full-set-configuration-perf}, we obtained substantial performance improvements 
for JavaScriptCore (JSC) on the Ostrich, Octane and Sunspider benchmark sets, indicating that the default configuration
of JSC leaves room for optimisation.
This is not the case for V8, where we did not find significant improvements for any of our benchmark sets,
which suggests that the default parameter values are already well optimised.
This may seem disappointing, but needs to be viewed in light of the fact that compiler developers test against these
same benchmarks, and have much incentive, through constant competition, to be successful in their efforts to find the best configurations.
It is therefore remarkable that we achieved sizeable performance gains for JSC, even on the SunSpider benchmark developed by the WebKit team (as noted earlier, WebKit uses the JSC engine).

\begin{table*}
    \begin{center}
        \begin{tabular}{l@{\hskip 2em}rrrr@{\hskip 1em}rrrr}
                                                    & \multicolumn{6}{c}{PAR10 [CPU s]} \\
        Instance set                    & JSC default & JSC configured & rel. impr. [\%] & V8 default    & V8 configured & rel. impr. [\%]            \\
        \midrule                                                                                                                                      
        Octane 2.0                       & 1.653       & 1.556          &  5.89\%      & 1.324         & 1.322         & 0.18\%                     \\
        Sunspider 1.0.2                  & 4.546       & 4.010          & 11.79\%      & 3.058         & 3.056         & 0.06\%                     \\
        Kraken                           & 1.214       & 1.205          &  0.72\%      & 0.650         & 0.650         & 0.03\%                     \\
        Ostrich                          & 9.739       & 9.263          & 4.88\%       & 7.109         & 7.062         & 0.66\%                     \\

        \end{tabular}
        \caption{
            Validation results using 100 runs per problem instance for each of our 4 configuration scenarios using complete instance sets.
            We give results for the JavaScriptCore and V8 default configurations, as well as for the best configuration
            obtained by SMAC (as identified by training performance).
            We note that the configurations found for JSC exhibit significant
            performance improvements over the entire instance set, while those
            for V8 show only marginal improvement over the defaults.
        }
        \label{tab:full-set-configuration-perf}
    \end{center}
\end{table*}

\subsection{Results on Individual Benchmark Instances}

When configuring the JavaScript engine parameters for individual instances from
our benchmark sets, we obtain much greater improvements than for the complete
sets. We selected the five most promising individual instances for the
experiments in this section to keep the resource requirements moderate.
We chose the instances based on where we observed performance improvements in
the experiments that optimised the configuration across the entire benchmark
sets.

\begin{figure*}
    \begin{center}
        \subfigure[JSC - Octane - PDFjs]{
            \includegraphics[width=0.4\textwidth]{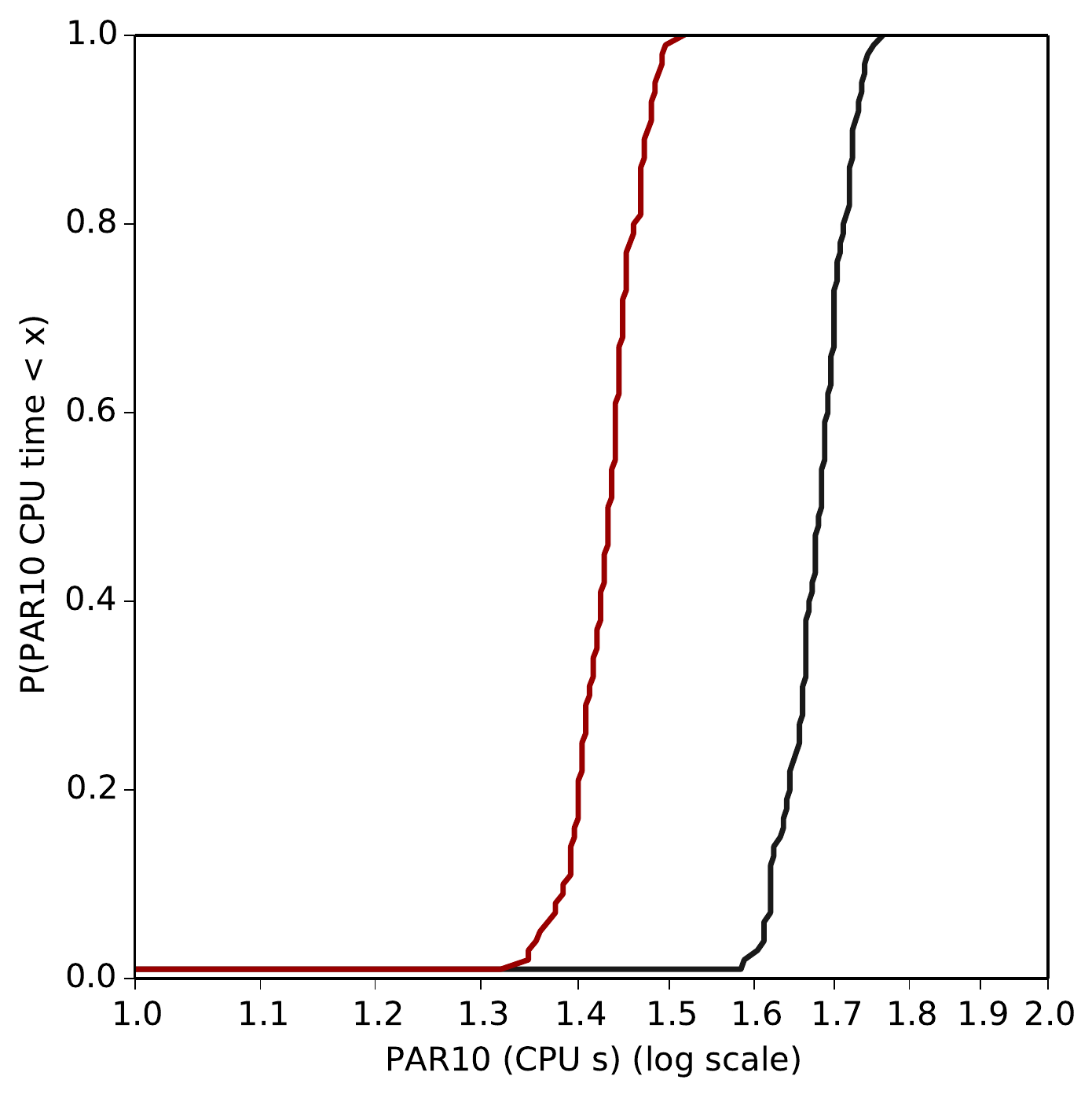}
            \label{fig:jsc-octane-pdfjs-rtd}
        }
        \subfigure[JSC - Octane - PDFjs]{
            \includegraphics[width=0.4\textwidth]{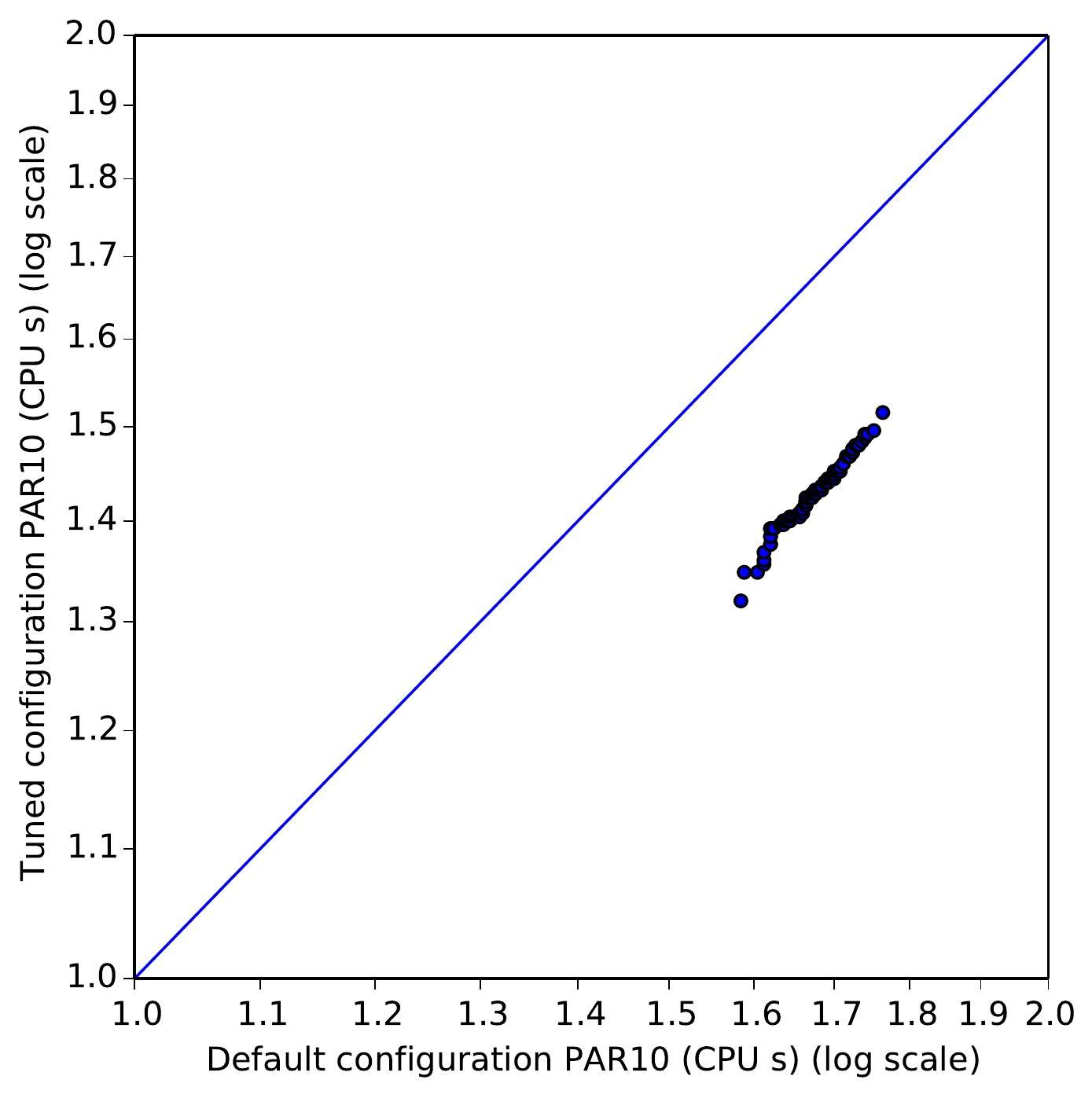}
            \label{fig:jsc-octane-pdfjs-scatter}
        }
        \\
        \subfigure[JSC - Octane - Splay]{
            \includegraphics[width=0.4\textwidth]{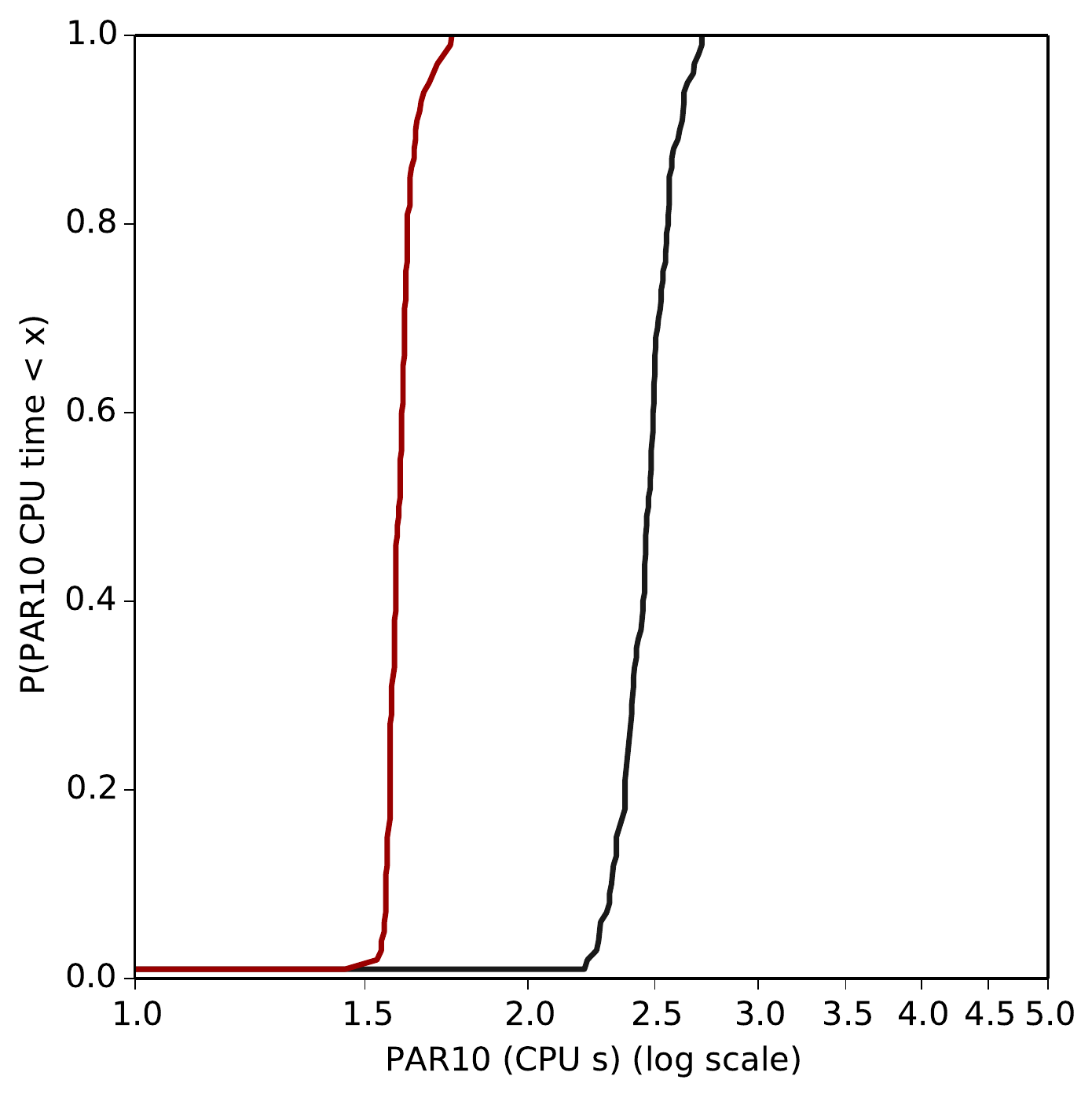}
            \label{fig:jsc-octane-splay-rtd}
        }
        \subfigure[JSC - Octane - Splay]{
            \includegraphics[width=0.4\textwidth]{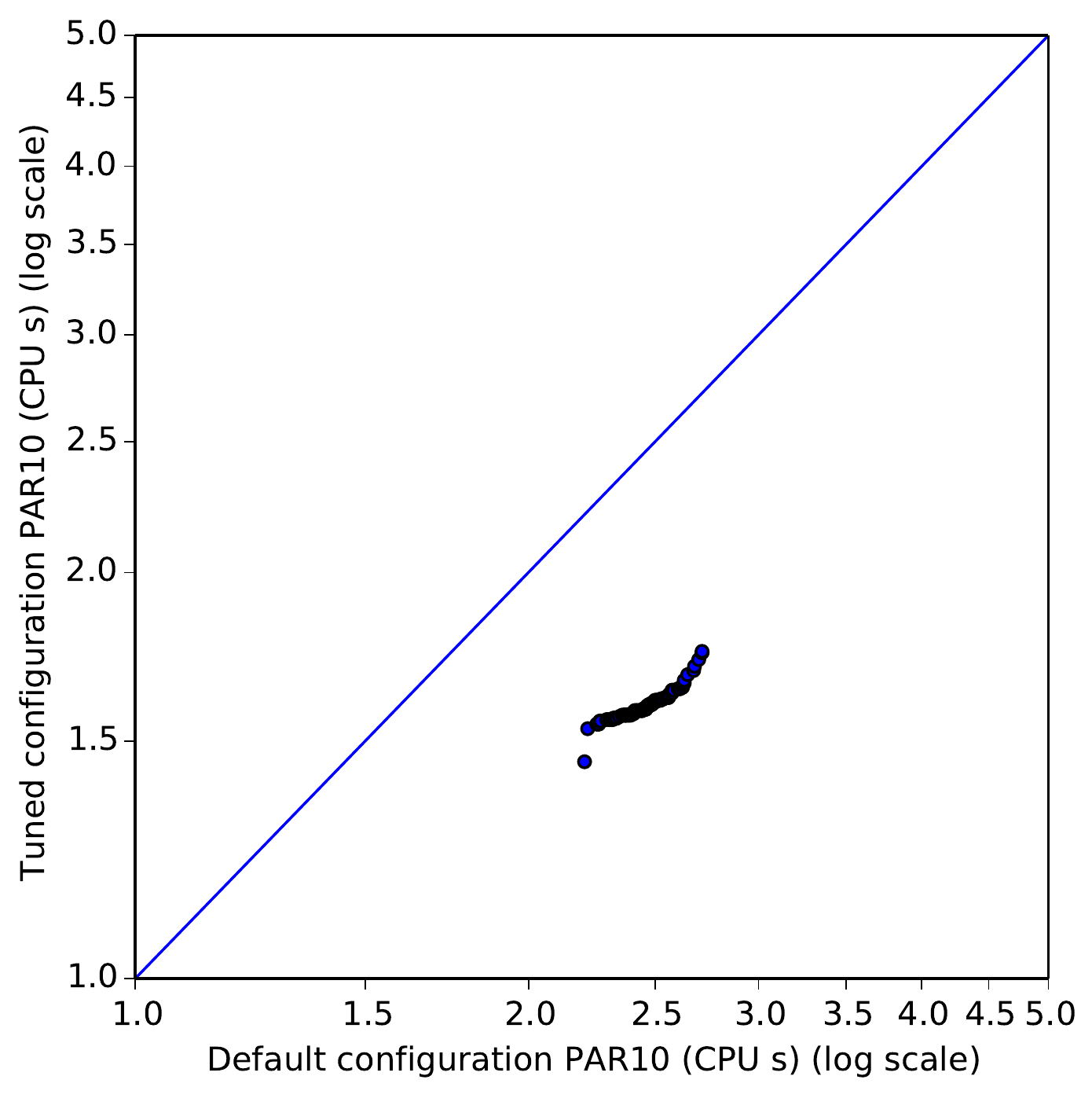}
            \label{fig:jsc-octane-splay-scatter}
        }
        \caption{
            For the Octane Splay and PDFjs invidual-instance configuration scenarios, we
            show empirical CDFs of runtime for 100 runs on the respective problem instance, along
            with scatter plots vs. the default configuration.
        }
        \label{fig:octane-rtd-scatter}
    \end{center}
\end{figure*}

\begin{figure*}
    \begin{center}
        \subfigure[JSC - Ostrich - Sparse LA]{
            \includegraphics[width=0.4\textwidth]{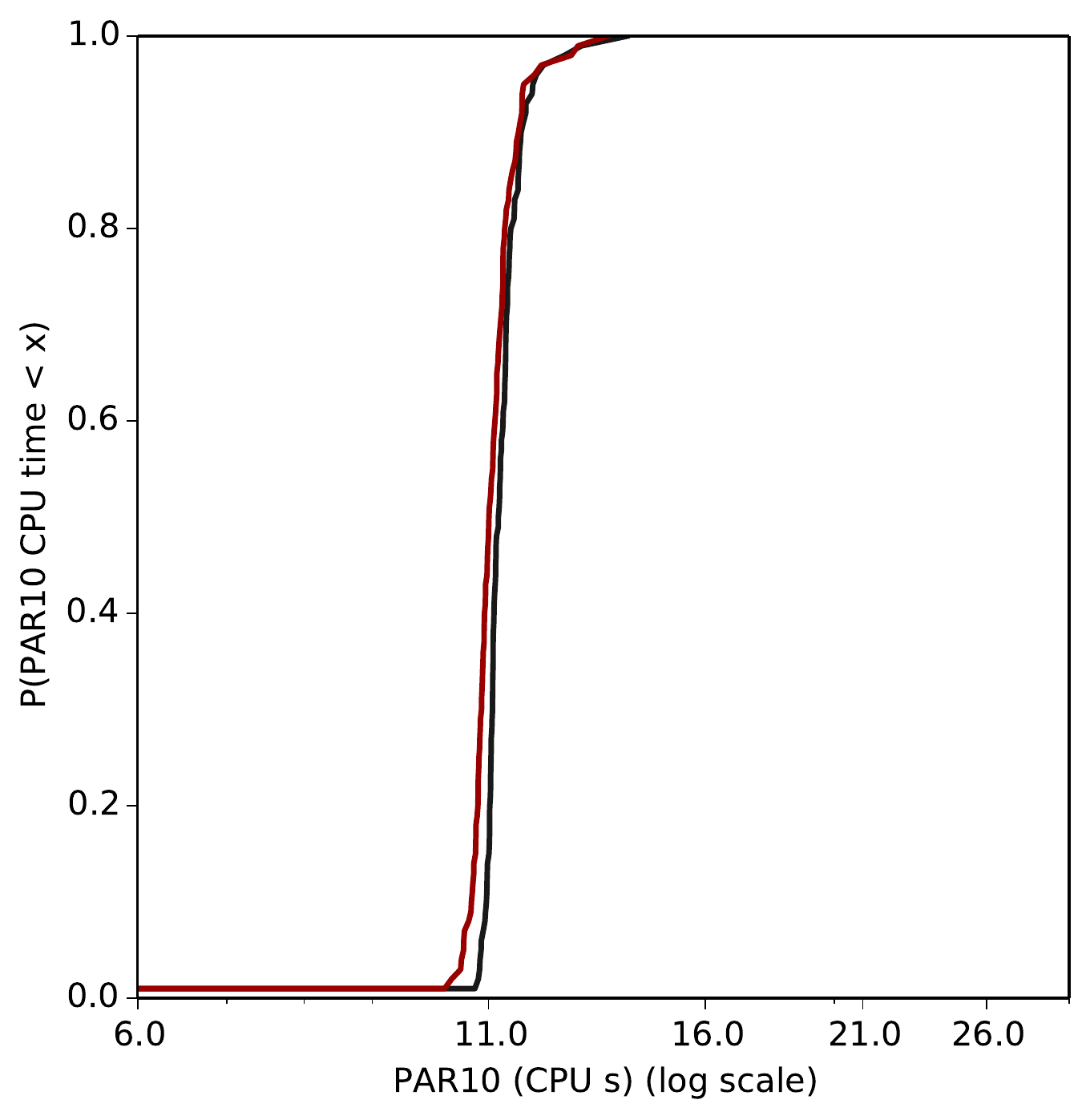}
            \label{fig:jsc-ostrich-sparse-rtd}
        }
        \subfigure[JSC - Ostrich - Sparse LA]{
            \includegraphics[width=0.4\textwidth]{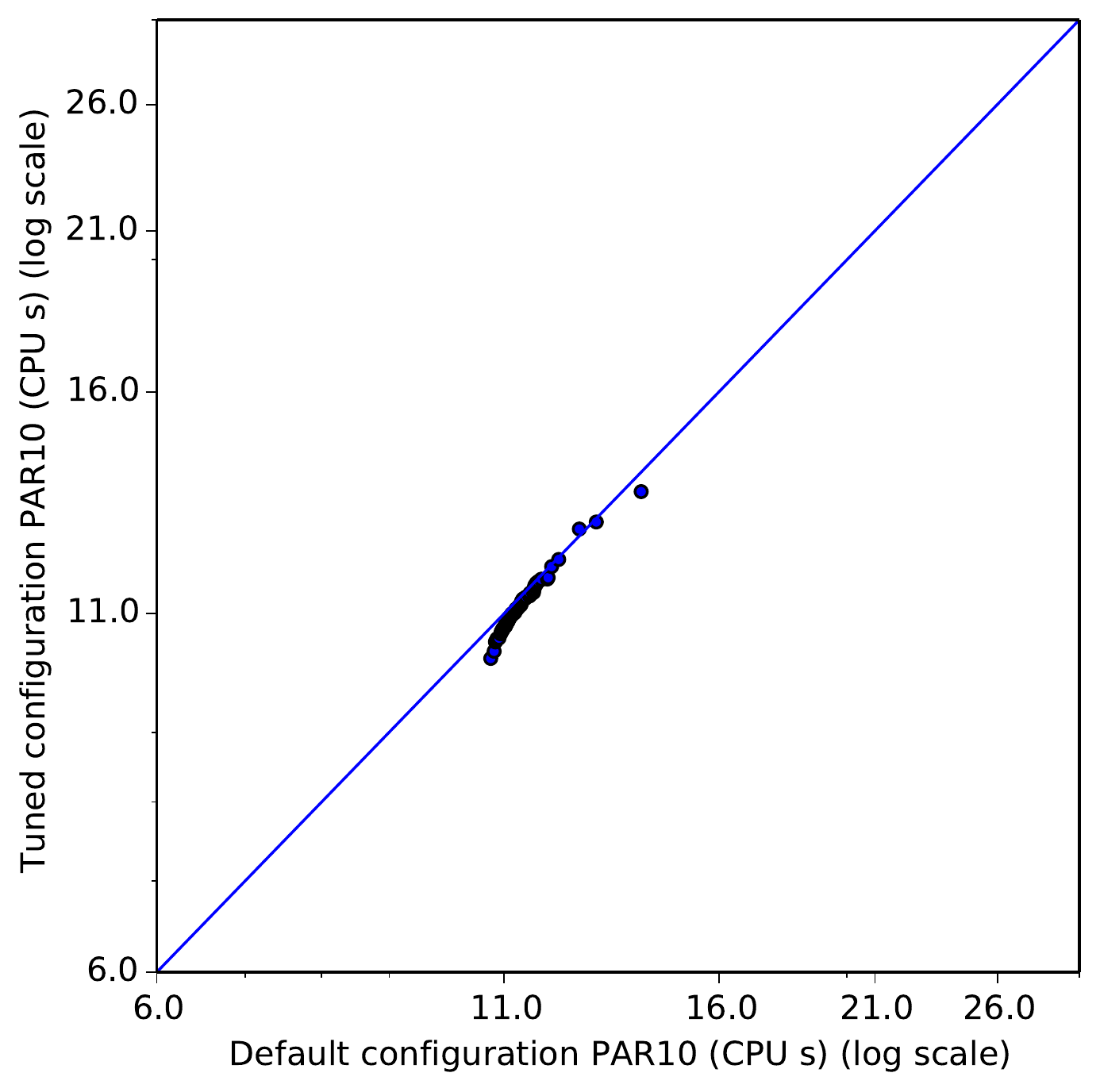}
            \label{fig:jsc-ostrich-sparse-scatter}
        }
        \\
        \subfigure[V8 - Ostrich - Sparse LA]{
            \includegraphics[width=0.4\textwidth]{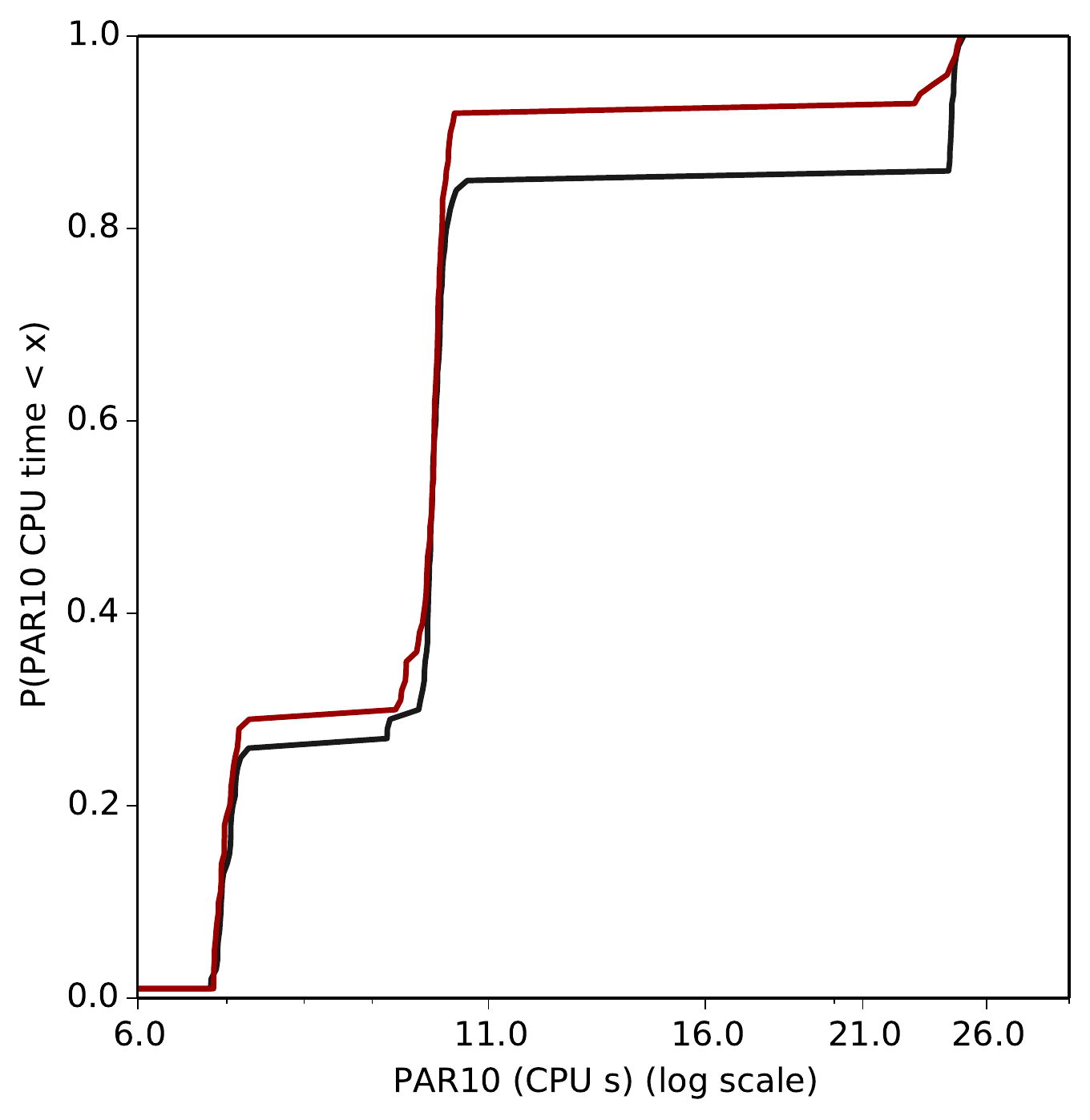}
            \label{fig:v8-ostrich-sparse-rtd}
        }
        \subfigure[V8 - Ostrich - Sparse LA]{
            \includegraphics[width=0.4\textwidth]{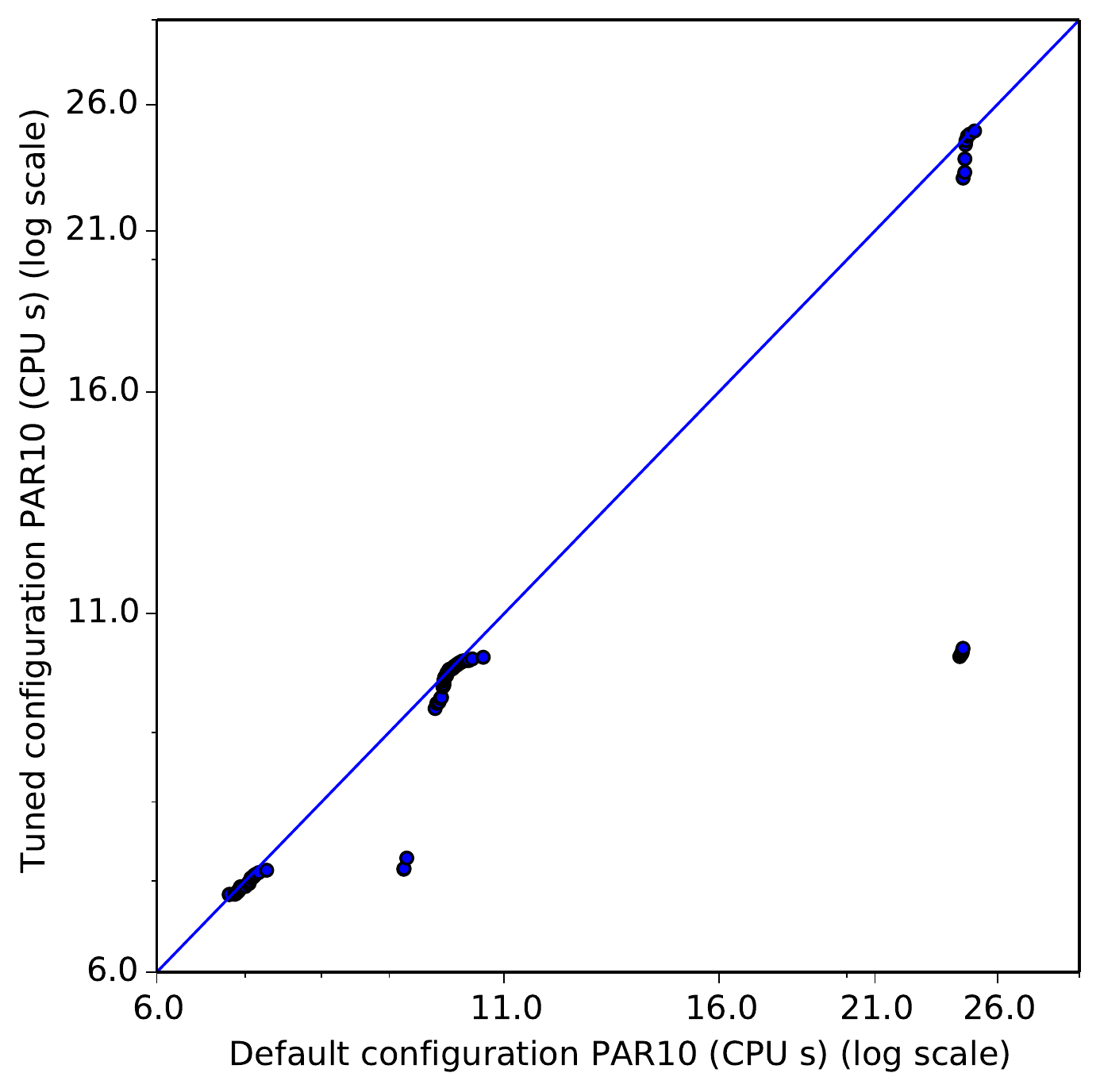}
            \label{fig:v8-ostrich-sparse-scatter}
        }
        \caption{
            Considering the Ostrich Sparse Linear Algebra individual-instance configuration scenario, we
            show empirical cumulative distribution functions (CDFs) of runtime for 100 runs on the respective problem instance, along
            with scatter plots vs. the default configuration. The CDFs show the
            probability that a run will complete within a certain amount of time
            as a function of the time, as observed empirically. That is, a
            finished run of an instance at a particular time increases the
            probability.
        }
        \label{fig:ostrich-sparse-rtd-scatter}
    \end{center}
\end{figure*}

Three of these instances are taken from the Ostrich set: graph-traversal, sparse-linear-algebra, and structured-grid,
and two instances stem from the Octane set: PDFjs and Splay.
Results from these experiments are shown in Table~\ref{tab:individual-instance-configuration-perf}, and we show additional
empirical cumulative distribution functions of running time and scatter plots for the default \emph{vs} optimised configuration in Figure~\ref{fig:ostrich-sparse-rtd-scatter} and Figure~\ref{fig:octane-rtd-scatter}.
On Ostrich graph-traversal or structured-grid, not shown in the table and figures, we have not obtained 
significant performance improvements for either of the two engines.

\begin{table*}
    \begin{center}
        \begin{tabular}{l@{\hskip 2em}rrrr@{\hskip 1em}rrrr}
                                                    & \multicolumn{6}{c}{PAR10 [CPU s]} \\
        Instance set                     & JSC default & JSC configured & rel. impr. [\%] & V8 default    & V8 configured & rel. impr. [\%]             \\
        \midrule                                                                                                                                        
        Ostrich sparse-linear-algebra   & 11.290      & 11.107         &  1.62\%        & 11.401        & 10.246        & 10.13\%                     \\
        Octane splay                    &  2.467      &  1.598         & 35.23\%        & 1.127         & 1.093         &  2.95\%                     \\
        Octane pdfjs                    &  1.679      &  1.431         & 14.76\%        & 1.654         & 1.645         &  0.57\%                     \\

        \end{tabular}
        \caption{
            Validation results using 100 runs per problem instance for 3 configuration scenarios using a single problem instance,
            one from our Ostrich set (Sparse Linear Algebra) and two from the Octane set (Splay and PDFjs).
            We omit two other experiments on Ostrich instances (Graph Traversal and Structured Grid), where neither compiler showed
            any improvement after configuration.
            We give results for the JavaScriptCore and V8 default configurations, as well as for the best configuration
            obtained by SMAC (as identified by training performance).
        }
        \label{tab:individual-instance-configuration-perf}
    \end{center}
\end{table*}

Overall, the performance improvements on these individual-instance configuration scenarios are surprisingly pronounced. JavaScriptCore achieves a relative performance improvement of 35.23\%
over the default configuration on the Octane Splay benchmark, and of 14.76\% on Octane PDFjs.
For V8, we observed a 10.13\% improvement over the default on Ostrich sparse-linear-algebra.

\todo{HH}{Add some discussion of the RTDs (mention randomisation in the code, measurement noise) - make it clear why we even need multiple runs on the same instance for evaluation. I don't understand why the instances shown in Figure 1 were chosen. I'd show detailed results for those mentioned above in the text.)}

Overall, these results are remarkable as even new code optimisation methods often only
result in performance improvements by single-digit percentages. We hypothesize that there are some specific aspects of these problem instances
which differ sufficiently from the other instances in their respective benchmark sets, that these configurations cannot
be successfully be used across those entire sets, but are very effective on the individual instance in question. We present
some preliminary results towards identifying the source of these improvements in the following.

\subsection{Time to Find Improving Configurations}

Even when considering the remarkable performance improvements seen in our individual-instance configuration
experiments, there may be some concern about the time required to find these improving configurations,
given that we used 25 independent SMAC runs of 1 CPU day to achieve these.

Upon further investigation, in all of our individual-instance configuration scenarios, the final optimised configuration was
found in less than 3 CPU hours of runtime, with initial improvements over the default configuration typically
found in less than 5 CPU minutes.
Longer runtimes are required for the complete instance set configuration scenarios, but even in those
cases, the final configuration was found in less than 6 CPU hours, with initial improving configurations
typically being found in less than 1 CPU hour.

In practice, a much smaller configuration budget would be sufficient to achieve
qualitatively similar results. In fact, we observed the first improvements after only a
few minutes of configuration.

\subsection{Changed Parameter Values}

In order to better understand the source of our individual-instance performance improvements, we empirically analysed the parameters changed from their default values using \emph{ablation analysis}~\cite{fawcettHoos2015}. 
This approach has been previously used successfully to assess the importance of parameter changes observed in applications of 
automated algorithm configuration techniques to propositional satisfiability, mixed-integer programming and AI-planning problems.
Ablation analysis greedily constructs a path through the parameter configuration space from
the default to a given target configuration, selecting at each stage the single parameter modification resulting
in the greatest performance improvement. The order of the resulting modifications reflects the relative contribution to the overall performance improvements obtained by the configuration process, where later changes may occasionally achieve bigger improvements that would not have been possible before earlier modifications to the default configuration.
The three parameter modifications resulting in the greatest performance improvement for the Octane Splay and PDFjs
instances are shown in Table~\ref{tab:jsc-octane-splay-ablation} and Table~\ref{tab:jsc-octane-pdfjs-ablation}, respectively.

For JavaScriptCore on Octane Splay, the parameter changes that
achieved the most significant improvements are related to object tracking and garbage collection.
For the Octane PDFjs benchmark instance, the configuration process resulted in modifications to various parameters controlling
memory management and the aggressiveness of the code optimisation. We note that numberOfGCMarkers is
important in both cases, where the value is changed to 1 from a default of 7.
This parameter controls the amount of parallelism in the garbage collector. Here,
reduced parallelism avoids overhead and improves overall performance.

While the portion of the relative improvement indicated in the tables is approximate due to the
nature of the ablation analysis procedure, it appears that in both cases, over 90\% of the observed
relative improvement can be explained by the modification of the three parameters shown. This is
consistent with previous results using ablation analysis, where in many scenarios, the vast majority of the improvement
was observed to be achieved by modifying a small set of parameters.
Of course, identifying these parameters in \emph{post hoc} ablation analysis is much easier than determining them
within the configuration process that gave rise to the optimised configurations thus analysed.

\begin{table*}
    \begin{center}
        \begin{tabular}{l@{\hskip 2em}r@{\hskip 2em}rrr}
        distance from default & parameter modified              & from      & to       & approx. portion of rel. impr. \\
        \midrule
        1                     & numberOfGCMarkers               & 7         & 1        &  38\% \\
        2                     & minCopiedBlockUtilization       & 0.9       & 0.196    & 47\% \\
        3                     & collectionTimerMaxPercentCPU    & 0.05      & 0.292    & 6\% \\
        \end{tabular}
        \caption{
            Parameters modified from the respective default settings for JavaScriptCore in order to achieve the three highest
            marginal performance gains on the Octane 2.0 instance ``Splay'', as determined by ablation analysis.
            Reported marginal improvement is only approximate, as the ablation analysis procedure is not performing the full 100
            runs per instance validation as in our configuration experiments.
        }
        \label{tab:jsc-octane-splay-ablation}
    \end{center}
\end{table*}

\begin{table*}
    \begin{center}
        \begin{tabular}{l@{\hskip 2em}r@{\hskip 2em}rrr}
        distance from default & parameter modified              & from      & to       & approx. portion of rel. impr. \\
        \midrule
        1                     & likelyToTakeSlowCaseMinimumCount& 20        & 56       & 41\% \\
        2                     & numberOfGCMarkers               & 7         & 1        & 40\% \\
        3                     & forceDFGCodeBlockLiveness       & false     & true     & 16\% \\
        \end{tabular}
        \caption{
            Parameters modified from the respective default settings for JavaScriptCore in order to achieve the three highest
            marginal performance gains on the Octane 2.0 instance ``pdfjs'', as determined by ablation analysis.
            Reported marginal improvement is only approximate, as the ablation analysis procedure is not performing the full 100
            runs per instance validation as in our configuration experiments.
        }
        \label{tab:jsc-octane-pdfjs-ablation}
    \end{center}
\end{table*}

\subsection{Performance under Different Loads}

Modern computers have multiple processors, with multiple CPU cores each, and it is desirable to run
multiple processes simultaneously in order to take full advantage of the processing power thus provided.
However, other factors, such as shared caches, memory bandwidth and the I/O
subsystem can affect performance negatively, if too many processes are vying for
resources.

In order to investigate to which extent such factors may impact our experimental setup, we ran
different configurations of workloads. First, we utilized all 32 cores of the
machine used for our experiments by running 32 benchmark experiments in parallel. Second,
we ran only 8 experiments in parallel, leaving the remaining cores for operating system
processes.

\todo{HH}{I've reworded, but still find it unclear what was done. Was it the case that in the first case, 32 runs of the same benchmark instance were done in parallel? If so, we should state it clearly.}

The results show that there are significant differences. The graph-traversal
instance of the Ostrich benchmark set requires a large amount of memory 
and sufficient memory bandwidth. With the machine fully loaded, we observe that
we easily find a parameter configuration that performs better than the default.
On the lightly loaded machine we are unable to do so, and the benchmark runs
significantly faster than on the fully loaded machine, even with the improved
configuration. This clearly indicates a memory bottleneck that can be mitigated
through configuration.

The default configuration of JavaScriptCore performs well on the SunSpider,
Kraken and Octane benchmarks on the fully-loaded machine, and we were unable to
find a better configuration of parameter settings. On the lightly loaded machine,
on the other hand, we did find better configurations for SunSpider and Octane. This
may indicate that the JavaScriptCore default configuration is optimised for a
highly-loaded machine, which is unlikely, when the engine is used inside a
browser on a user's desktop or laptop machine.

The fact that JavaScriptCore  and V8 and exhibit different behaviour with respect to
how easy it is to improve on their default configurations on machines with
different load suggests that the benchmarking and tuning the respective development teams
perform may use different experimental setups.

This result shows that the optimisation of compiler flags should be done not only
for the machine that the code will be run on, but also for the expected load on
that machine -- configuring for a lightly loaded machine will yield different
results than configuring for a heavily loaded one. Furthermore, there is much promise
in switching between different configurations based on machine load.

\begin{table*}
    \begin{center}
        \scriptsize
        \begin{tabular}{l@{\hskip 2em}r@{\hskip 1em}rrrrrrrrrr}
                                        &           & \multicolumn{10}{c}{PAR10,s (ID) at rank} \\
Experiment                  & Default   & 1         & 2             & 3         & 4         & 5          & 6         & 7          & 8          & 9         & 10 \\
\midrule
JSC (32) Octane PDFjs 1     & 2.163     & 2.009 (11) & 2.020 (14)   & 2.037 (20)& 2.037 ( 9)& 2.042 (22) & 2.046 ( 3)& 2.047 (12) & 2.052 (15) & 2.060 ( 5)& 2.074 (18) \\
JSC (32) Octane PDFjs 2     & 2.976     & 2.890 ( 4) & 2.923 ( 1)   & 2.929 ( 2)& 2.963 ( 5)& 2.971 ( 7) & 2.987 ( 3)& 2.996 ( 9) & 3.037 (11) & 3.041 ( 6)& 3.060 (15) \\
JSC (32) Octane PDFjs 3     & 2.875     & 1.996 (14) & 2.005 (11)   & 2.035 (24)& 2.038 (20)& 2.043 (15) & 2.045 (22)& 2.053 (12) & 2.067 ( 5) & 2.068 ( 9)& 2.071 (16) \\
\midrule
JSC (8) Octane PDFjs 1     &  1.691     & 1.422 (11) & 1.434 (14)   & 1.460 (22)& 1.471 (24)& 1.473 (2)  & 1.477 (5) & 1.489 (20) & 1.513 (3)  & 1.555 (16)& 1.556 (18) \\
JSC (8) Octane PDFjs 2     &  1.687     & 1.426 (11) & 1.431 (14)   & 1.463 (22)& 1.466 (24)& 1.470 (2)  & 1.471 (5) & 1.479 (20) & 1.516 (3)  & 1.555 (16)& 1.558 (18) \\
JSC (8) Octane PDFjs 3     &  1.691     & 1.416 (11) & 1.426 (14)   & 1.459 (22)& 1.462 (24)& 1.471 (5)  & 1.471 (2) & 1.483 (20) & 1.525 (3)  & 1.548 (18)& 1.556 (16) \\

        \end{tabular}
        \caption{
            Using the Octane pdfjs problem instance, we performed 100 independent runs of the 25 SMAC configurations for JSC, as well as the JSC default configuration.
            This was repeated 3 times with the same random seeds, first allowing 32 simultaneous runs and again allowing 8 simultaneous runs.
            We give the PAR10 score for the default configurations, as well as for the 10 best configurations by validation score in each experiment (along with the configuration ID for each).
            The configuration ID for the ``best training'' configuration of JSC on this instance is 14.
            It is clear that the best configurations are quite different in the case of 32 simultaneous runs, even with a fixed instance and seeds.
            As this variability disappears in the case of 8 simultaneous runs, we attribute it to noise from the load (and subsequent cache contention, etc.).
        }
        \label{tab:machine-load-stability}
    \end{center}
\end{table*}

\section{Conclusions}

JavaScript is ubiquitous in the modern world wide web and increasingly spreading
into other areas that have traditionally been dominated by other programming languages. It is
used client-side in web browsers as well as server-side in backend
applications. Performance increasingly matters in practical JavaScript, as
applications grow in size and complexity.

In part, the success of JavaScript has been due to the availability of
highly optimised compilers that produce efficient code that can be executed
with minimal overhead. Just-in-time compilation and dynamic optimisations
further increase the performance of the code.
However, contemporary compilers have a large number of parameters, most of which
are only poorly documented. While the default configuration of these parameters
provides good performance in most cases, the parameter values need to be
optimised for the application at hand to get the best performance in all cases.
Exploring this huge and complex parameter space is a daunting task.

We apply a state-of-the-art, general-purpose automated configuration procedure
with an excellent track record in applications in machine learning and combinatorial optimisation 
to the problem of
finding the best parameter configuration for JavaScript engines for a particular
set of problem instances. Sequential model-based optimisation leverages state-of-the-art
techniques from statistics, optimisation and machine learning to efficiently and automatically explore the
parameter space of an algorithm and to home in on promising configurations quickly.

Our experimental evaluation shows that notable performance improvements can be achieved
through automated configuration. Specifically, we demonstrate that the
performance of JavaScriptCore can be substantially improved on 3 out of 4
heterogeneous benchmark sets in common use for JavaScript compiler benchmarking.
We also show that JavaScriptCore (and to a lesser extent V8) can be specialised
to obtain runtime gains of up to 35\% on tasks such as PDF rendering.
This is particularly significant as we are optimising code that is run millions
of times. In contrast, algorithm configuration for combinatorial optimisation
problems considers the different setting where each problem instance needs to be
solved only once.

We believe that our results are promising and believe that our approach enables
many interesting applications and follow-up work.
We are currently planning additional work including a broader set of experiments,
additional analysis of the parameter space structure, a deeper investigation
into the effect of machine load on runtime performance and configuration, and an
investigation of the transferability of these configuration results to machines
other than those used for training.

\subsection*{Acknowledgements}
Part of this research was supported by a Microsoft Azure for Research grant.
HH also acknowledges support through an NSERC Discovery Grant.

\clearpage

\bibliography{\jobname}
\bibliographystyle{aaai}

\end{document}